\documentclass[aps,prmaterials,reprint,superscriptaddress]{revtex4-2}

\pdfoutput=1
\usepackage{graphicx,color}
\usepackage[center]{subfigure}
\usepackage{mathtools}
\usepackage{bm}
\usepackage{esvect}
\usepackage{enumitem}
\usepackage{wrapfig}
\usepackage{ulem}
\usepackage[mathscr]{euscript}

\newcommand \fd[2] {\frac{\delta #1}{\delta #2}}
\newcommand \pt[2] {\frac{\partial #1}{\partial #2}}

\newcommand \bedm {\begin{dmath}}
\newcommand \eedm {\end{dmath}}
\newcommand \be {\begin{eqnarray}}
\newcommand \ee {\end{eqnarray}}
\newcommand \ben {\begin{eqnarray}}
\newcommand \een {\end{eqnarray}}
\newcommand \Lap {\nabla^2}

\newcommand \Grad {\vec{\nabla}}
\newcommand \nline {\nonumber \\}

\newcommand \etal {{\it et al.\ }}
\newcommand \qeq {q^{\rm eq}_0}

\begin{document}

\title{Influence of dislocations in multilayer graphene stacks: A phase field crystal study}

\author{K. R. Elder}
\affiliation{Department of Physics, Oakland University, Rochester, Michigan 48309, USA}

\author{Zhi-Feng Huang}
\affiliation{Department of Physics and Astronomy, Wayne State University, Detroit, Michigan 48201, USA}

\author{T. Ala-Nissila}
\affiliation{QTF Centre of Excellence, Department of Applied Physics, 
Aalto University, P.O. Box 15600, FI-00076 Aalto, Espoo, Finland}
\affiliation{Interdisciplinary Centre for Mathematical Modelling, Department of Mathematical Sciences, 
Loughborough University, Loughborough, Leicestershire LE11 3TU, UK}

\begin{abstract}
In this work the influence of $5|7$ dislocations 
in multiplayer
graphene stacks (up to six layers) is examined.  The study is conducted
through a recently developed Phase Field Crystal (PFC) model for multilayer
systems incorporating out-of-plane deformations and
parameterized to match to density functional theory calculations for graphene
bilayers and other systems. The specific configuration considered consists of one
monolayer containing four $5|7$ dislocations (i.e., two dislocation dipoles) sandwiched in between perfect graphene layers.  The
study reveals how the strain field from the dislocations in the defected layer
leads to out-of-plane deformations that in turn cause deformations of
neighboring layers.  Quantitative predictions are made for the defect free energy 
of the multilayer stacks as compared to a defect-free system, which is shown to increase with the number 
of layers and system size.  Furthermore it is predicted that system defect energy saturates by roughly ten
sheets in the stack, indicating the range of defect influence across the multilayer.  Variations of stress field distribution and layer height profiles in different layer of the stack are also quantitatively identified.

\end{abstract}

\maketitle

\section{Introduction}

 Since the discovery of graphene (Gr) \cite{Novoselov2004}, monolayers of
 various two-dimensional (2D) materials have been the focus of many research
 studies due to their potential use in technological applications. Currently a vast
 collection of materials can exhibit stable (free-standing or supported) 2D structures, such as Gr and its various
 allotropes, silicene (Si), phosphorene (P), hexagonal boron nitride (hBN),
 MoS$_2$, NbSe$_2$, and SnTe, among many others. Some of them display an extraordinary variety of
 correlated magnetic and electronic states \cite{You21,Burch18}. Their range of
 physical properties can be further varied by combining them either laterally or vertically. Selected
 immiscible materials such as Gr and hBN can even coexist within a single 2D layer
 \cite{Pete19}, and in-plane heterostructures or lateral multijunctions of transition-metal dichalcogenides (TMDs) can be grown via edge epitaxy \cite{Zhang17,Huang22}. Also importantly, due to the dominant dispersive van der Waals
 (vdW) forces between layers, it is possible to stack them into coupled 2D
 multilayers (s2D) that can be either identical such as in multilayer Gr or form
 heterogeneous vdW structures from different materials such as Gr and hBN or different types of TMDs.
 
 Stacked layers of vdW materials can be further twisted or translated relative to each other. This often leads to exotic
 material patterns, phenomena, and properties. For example, twisted Gr bi- and
 multilayer s2D systems lead to
 complex ``flamingo" type Moir\'e patterns \cite{Ouyang2021} and generate novel
 correlated states of matter such as heavy fermions \cite{Vano21} and van Hove
 singularities \cite{Havener2014}. One of the most spectacular examples is
 that twisted Gr bilayers exhibit superconductivity at a “magic” twist angle
 ($\approx 1^\circ$) \cite{Guinea2018,Andrei2020}. The number of possible stacking
 combinations of 1D/2D/3D materials is enormously large and will remain an
 active field of research for a long time to come.

The focus of our work here is on multiple layers of Gr with one layer containing
$5|7$ defects.  In general such defected states and others (such as 
the superlattice Moir\'e patterns discussed above) can significantly 
alter electrical, thermal and mechanical properties in 2D materials 
and their stacking.
Defects, including dislocations, grain boundaries,
and impurities, often strongly influence material properties, such as thermal
resistance in polycrystal Gr \cite{Azizi2017} and hBN \cite{Dong2018}.   Gong \etal
\cite {Gong17} have shown that in a Gr bilayer, with one layer containing a
point defect, the barrier for defect migration is increased by 9 -- 14\%
compared to a single monolayer, while Choi \etal \cite{Choi2008} found that the
spin magnetic moment near a monovacancy in Gr is reduced when another Gr layer
is added. Recent experiments by Mohapatra \etal \cite{Mohapatra2021} have shown
that even wrinkles in few-layer Gr materials can increase the thermal
conductivity ($\kappa$) from 165 W/mK in pristine regions to 225 W/mK near
wrinkled regions.  Thus it is important to understand the impact 
of defects in these quasi-2D systems. 

In this paper the influence of $5|7$ dislocation defects 
in one layer of Gr in (or close to) the middle of a stack of Gr multilayers is considered.
Our study is based on a 2D phase field crystal model which allows for out-of-plane
deformations \cite{Elder21}.  In general PFC methods have an advantage over atomistic methods (e.g., molecular dynamics) in 
that the simulation timescale is controlled by diffusive time scales as opposed to vibrational 
scales.  In addition, PFC models are based on system free energies, not atomic potentials, so 
that entropy, for example, is naturally incorporated.
The out-of-plane PFC model has been
applied to study $5|7$ defects in a single Gr layer and predicted an energy
of 6.8 eV which is similar to that from quantum-mechanical density functional theory (DFT)
calculations of Chen and Chrzan \cite{Chen2011} (6.17 eV) and Yazyev and Louie
\cite{Yazyev10} (7.5 eV), and molecular dynamics calculations of Liu and Yakobson
\cite{Liu10} (5.0 eV). The model has also been used to identify the scaling
behavior of height fluctuations in both pinned and free-standing Gr monolayers
\cite{Enzo23,Enzo22} as well as Moir\'e patterns in Gr/hBN bilayers \cite{Elder23}.  
The results of these prior studies  
were consistent with prior experiments,
simulations and theoretical predictions. Furthermore the interaction between Gr
layers was parameterized to match the stacking energies and heights predicted by
DFT calculations of Zhou \etal \cite{Zhou15} as discussed in Sec.
\ref{sec:meth}.

The paper is structured as follows. In Sec. \ref{sec:meth} the methodology 
and model paramaterization used are presented for the study of 
multiple layers of stacked Gr.  In Sec. \ref{sec:disl} the influence 
of two dislocation dipoles (each containing two $5|7$ defects) within 
one layer surrounded by other pristine Gr layers is examined, with predictions of system energy, stress field distribution, and layer height profiles as well as their variations on the number of stacking layers and system size provided.  Finally in 
Sec. \ref{sec:dissum} a discussion and summary of the results are 
provided.

\section{Methodology}
\label{sec:meth}

Consider a stack of $M$ layers each of which is described by a 
2D dimensionless atomic number density variation field,
$n_i(x,y)$, and a height, $h_i(x,y)$.  As discussed in a prior publication 
\cite{Elder21} a free energy functional ${F}_i$ that 
is minimized by a periodic array of maxima in $n_i$ and allows for 
out-of-plane deformations can be written as
\be
\frac{{F}_i}{c_{\rm g}} = &&
\int d\vec{r} \left[
\frac{\Delta B}{2} n_i^2+\frac{B^x}{2} 
\left({\cal L}_w n_i\right)^2
+\frac{\tau}{3}n_i^3+
\right. \nline && \left. +
\frac{v}{4}n_i^4+ 
\frac{\kappa}{2}  \int d\vec{r}^{\,\prime} 
C(|\vec{r}-\vec{r}^{\,\prime}|) h_i(\vec{r}) h_i(\vec{r}^{\,\prime})\right],
\label{eq:ftot}
\ee
where 
\be
{\cal L}_w = 1+\nabla^2-(
h_x^2\partial_x^2
+h_y^2\partial_y^2
+2h_xh_y\partial_{x}\partial_y),
\ee  
with $h_x=\partial_x h_i$, $h_y=\partial_y h_i$, and $\nabla^{2}=\partial_x^2+\partial_y^2$. 
The last term in Eq.~(\ref{eq:ftot}) accounts for the bending energy of the plane and the parameter $\kappa$ is the bending energy coefficient.
The Fourier transform of kernel $C$ is set as $\hat{C}_k=-k^4$ for wave number $k<1/2$ and $\hat{C}_k=-10^4$ for $k>1/2$, to suppress short range fluctuations in the height. The results reported here were found to be insensitive to  
the choice of this $\hat{C}_k$ cutoff, as long as the cutoff was large enough to eliminate short range fluctuations in $h$ but short enough to not change the dislocation core configuration or energy. 
The parameters for graphene are set to $(c_{\rm g},\Delta B,B^x,\tau,v,\kappa,\bar{n}_i)
=(6.58{\rm eV},-0.15,0.874818,1,0.209726,0)$, where $\bar{n}_i$ is the average 
density value of $n_i$.  
For these parameters the lowest energy state is a 
2D honeycomb array of density maxima and the energy of 
a $5|7$ defect has been calculated to be $6.8$ eV.

For a bilayer system the direct coupling between layers occurs through
\begin{equation}
{F}^{\rm c}_{ij} {\rm =} \int d\vec{r}\left[
\sum_{l=1}^2V_0^{(l)} (\delta n_i\delta n_j)^l
+a_2 \left ( \Delta h_{ij} - \Delta h^0_{ij} \right )^2
\right],
\label{eq:Fc2}
\end{equation}
where $\Delta h_{ij} = h_i-h_j$, $\delta n_i=n_i-\bar{n}_i$ 
and  $\Delta h^0_{ij} = \Delta(1+\alpha\delta n_i\delta n_j)$.  
As discussed in Ref.~\cite{Elder21}, the coupling parameters 
$V_0^{(1)},V_0^{(2)}$, $\alpha_2$, $\Delta$, and $\alpha$ can be fit to the 
stacking energies and heights that were calculated by Zhou \etal \cite{Zhou15} 
using quantum-mechanical density functional theory (DFT) (more specifically the 
computationally expensive, but more accurate, ACFDT-RPA results as used
in this work). For a system of $M$ layers this can easily be extended 
by assuming each layer only interacts with neighbouring 
layers (via Eq. (\ref{eq:Fc2})).
The goal of this work is to determine the lowest energy states 
for a set of layers with one of them containing defects.  The free energy minimization 
is obtained by relaxational dynamics, i.e., 
\be
\pt{n_i}{t}&=&-\fd{{ F}_{\rm t}}{n_i}; \label{eq:ndt}\\
\pt{h_i}{t}&=&-\Gamma \fd{{ F}_{\rm t}}{h_i},
\label{eq:hdt}
\ee
where $i=1,2,...,M$, $\Gamma$ is the mobility for layer height, and ${ F}_{\rm t}$ is the sum of the free
energies of all layers and their interaction energy ($F^{\rm c}$), i.e., 
\be
{ F}_{\rm t} = \sum_{i=1}^M { F}_i + \sum_{i=1}^{M-1} { F}^{\rm c}_{i+1,i} = \sum_{i=1}^M { F}_i + F^{\rm c}.
\label{eq:f_tot}
\ee
Simulations were performed on the $xy$ plane with periodic boundary conditions, using the
semi-implicit method in Fourier space as described in Ref.~\cite{Nik2010} Appendix B 4. 
The average value of $n_i$ for each layer was fixed (i.e., the $k=0$ Fourier mode of $n_i$ was never 
changed).
For the largest system size studied here (i.e., six layers with each of size $138~{\rm nm} \times 69~{\rm nm}$), relaxation to a steady state could take considerable amount of computational time, on the order of one month in a single GPU.

In general a one-mode approximation for the density field, i.e., 
\be
n_i = \bar{n}_i + 2\phi \sum_{j=1}^3 \cos(\vec{G}_j\cdot \vec{r}),
\label{eq:onem}
\ee
was typically used as the initial state, where the reciprocal lattice vectors are
$\vec{G}_1=(-\sqrt{3}q_x/2,-q_y/2)$, $\vec{G}_2=(0,q_y)$, and
$\vec{G}_3=(\sqrt{3}q_x/2,-q_y/2)$.  In this approximation the values of $\phi$,
$q_x$ and $q_y$ that minimize the free energy are given by
\begin{equation}
\phi_{\rm eq} = \frac{3\bar{n}_iv+\tau+\sqrt{\tau^2-15v\Delta B -4 \bar{n}_iv(6\tau+9\bar{n}_iv)}}{15v},
\label{eq:phieq}
\end{equation}
and $q_x=q_y=1$ with $h_i(\vec{r})=0$.  To obtain a more accurate equilibrium state 
the value of $q_x=q_y$ was varied for a single unit cell and $n_i$ was 
relaxed via Eq. (\ref{eq:ndt}) until equilibrium was reached (with $h_i=0$).
In all simulations a 2D periodic box was used with grid spacing chosen to 
exactly fit the box, such that $\Delta x\approx 1/2$ and $\Delta y\approx 1/2$.
The simulation results gave equilibrium wave numbers of $q_x=q_y=\qeq=0.9976666$.

\subsection{Bilayer parameter fitting}

\begin{figure}[ht]
\vskip 0pt
\includegraphics[width=0.45\textwidth]{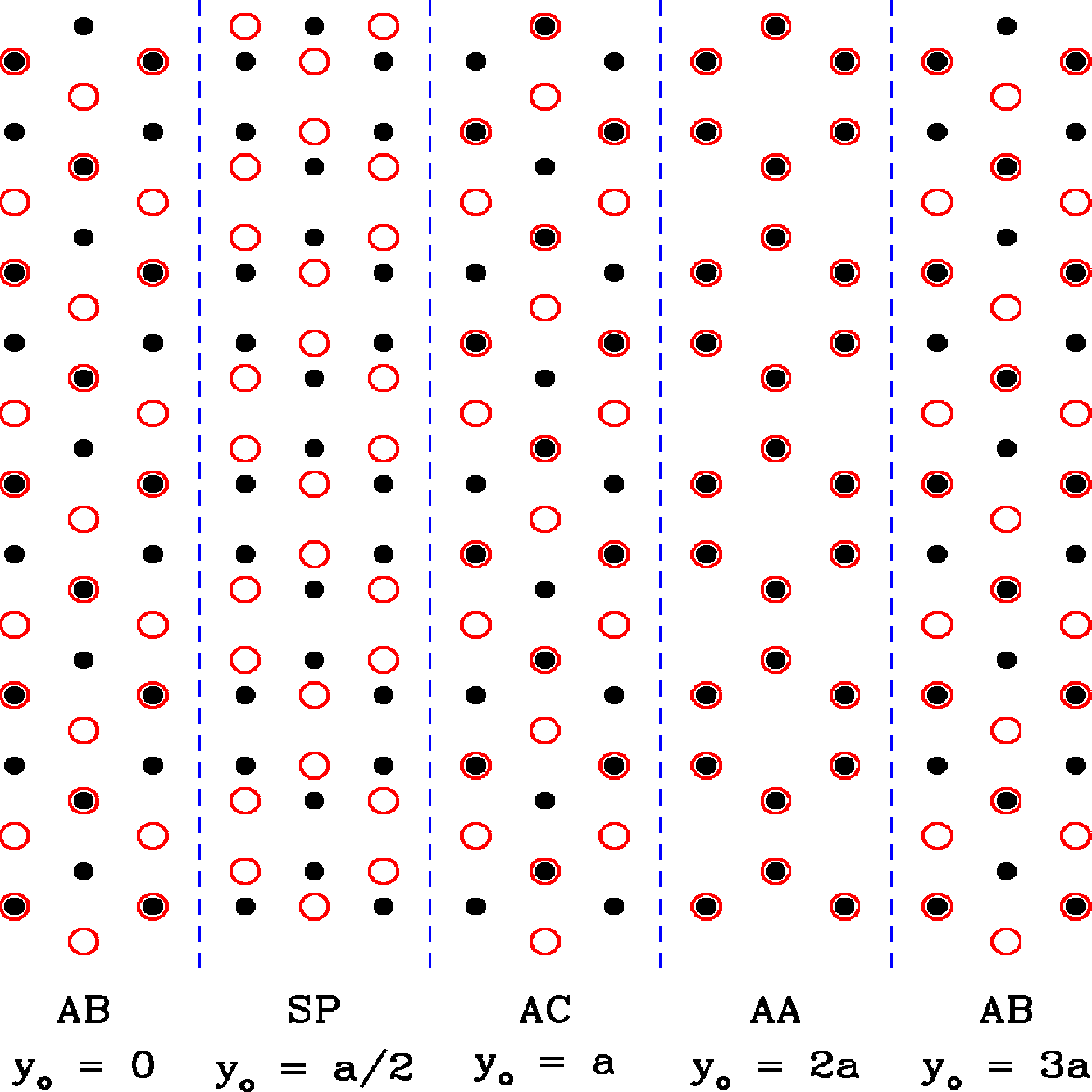}
\caption{Illustration of different stackings of two graphene layers with shifting distance $y_0$, 
where the black solid (red empty) points correspond to the bottom (top) layer.}
\label{fig:GGrN}
\end{figure}

In a prior publication \cite{Elder21} the parameters entering Eq.~(\ref{eq:Fc2})
were chosen to best fit the DFT calculations of Zhou \etal \cite{Zhou15} for
various bilayer stacking configurations, as depicted in Fig.~\ref{fig:GGrN}.
For simplicity the calculations were performed assuming that the densities could
be represented by only the lowest-order Fourier components needed to describe
the honeycomb symmetry (i.e., Eq.~(\ref{eq:onem})). However, numerical solutions
of the full PFC model naturally incorporate contributions from higher-order
modes. To account for this, simulations of the full model were performed in a
single unit cell starting from the initial condition with $h_i=0$ and $n_i$
described by Eq.~(\ref{eq:onem}) with $(\phi,q)=(\phi_{\rm eq},\qeq)$. The
system was first relaxed in an AB type stacking (see Fig.~\ref{fig:GGrN}), and
then one layer was shifted with respect to the other and the relative heights
($h_1$ and $h_2$) were allowed to relax via Eq.~(\ref{eq:hdt}) but not the
densities $n_i$.  Allowing the densities to vary in a non-equilibrium stacking
would lead to the system relaxing back into the state of AB stacking.   

\begin{figure}[ht]
\vskip 0pt
\includegraphics[width=0.45\textwidth]{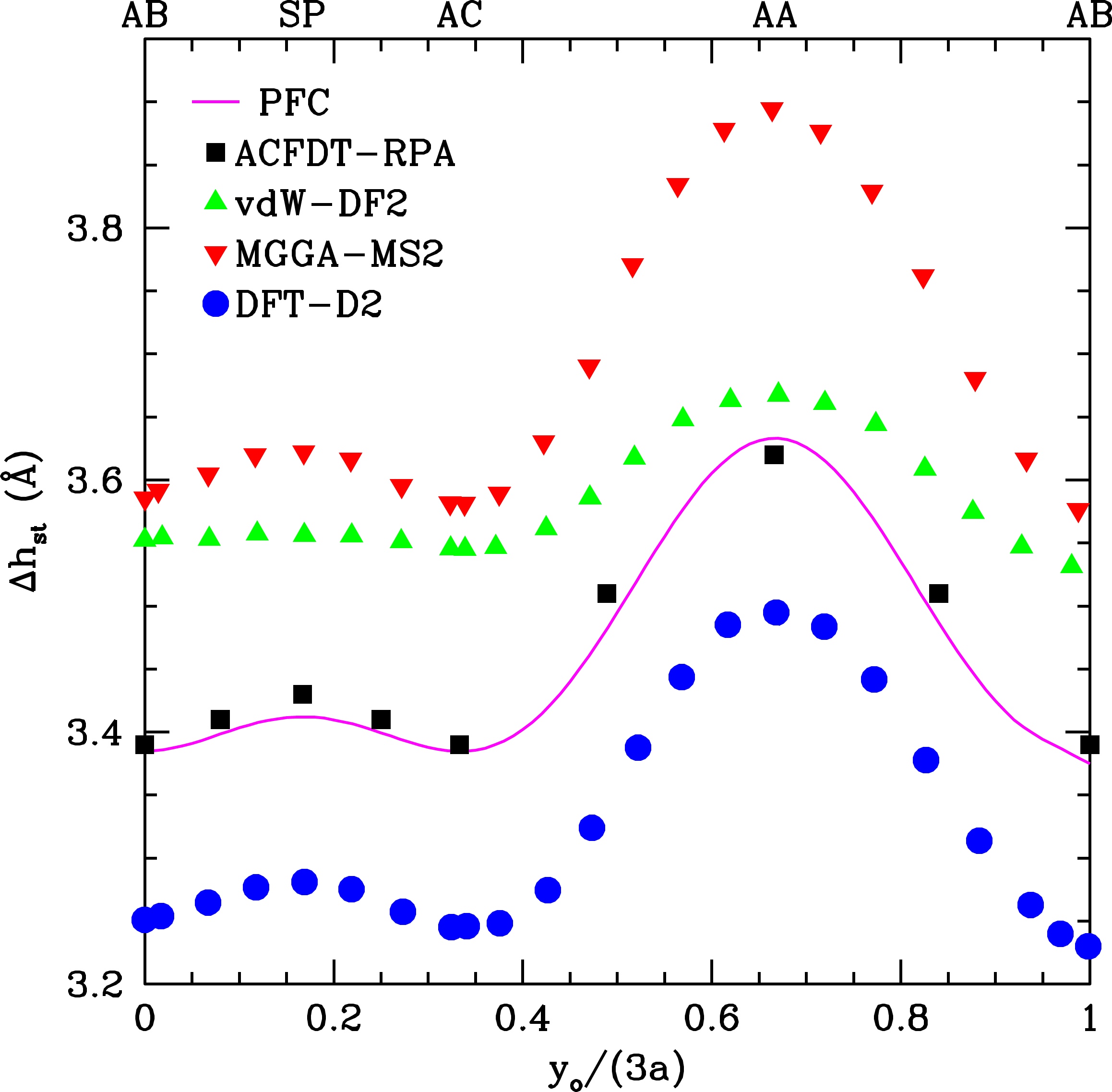}
\caption{Comparison of stacking height difference obtained from various DFT calculations 
\cite{Zhou15} with those from the PFC bilayer model. See Fig.~\ref{fig:GGrN} for illustration 
of the stackings labelled as AB, SP, AC, and AA.}
\label{fig:HggVc}
\end{figure}

\begin{figure}[ht]
\vskip 0pt
\includegraphics[width=0.45\textwidth]{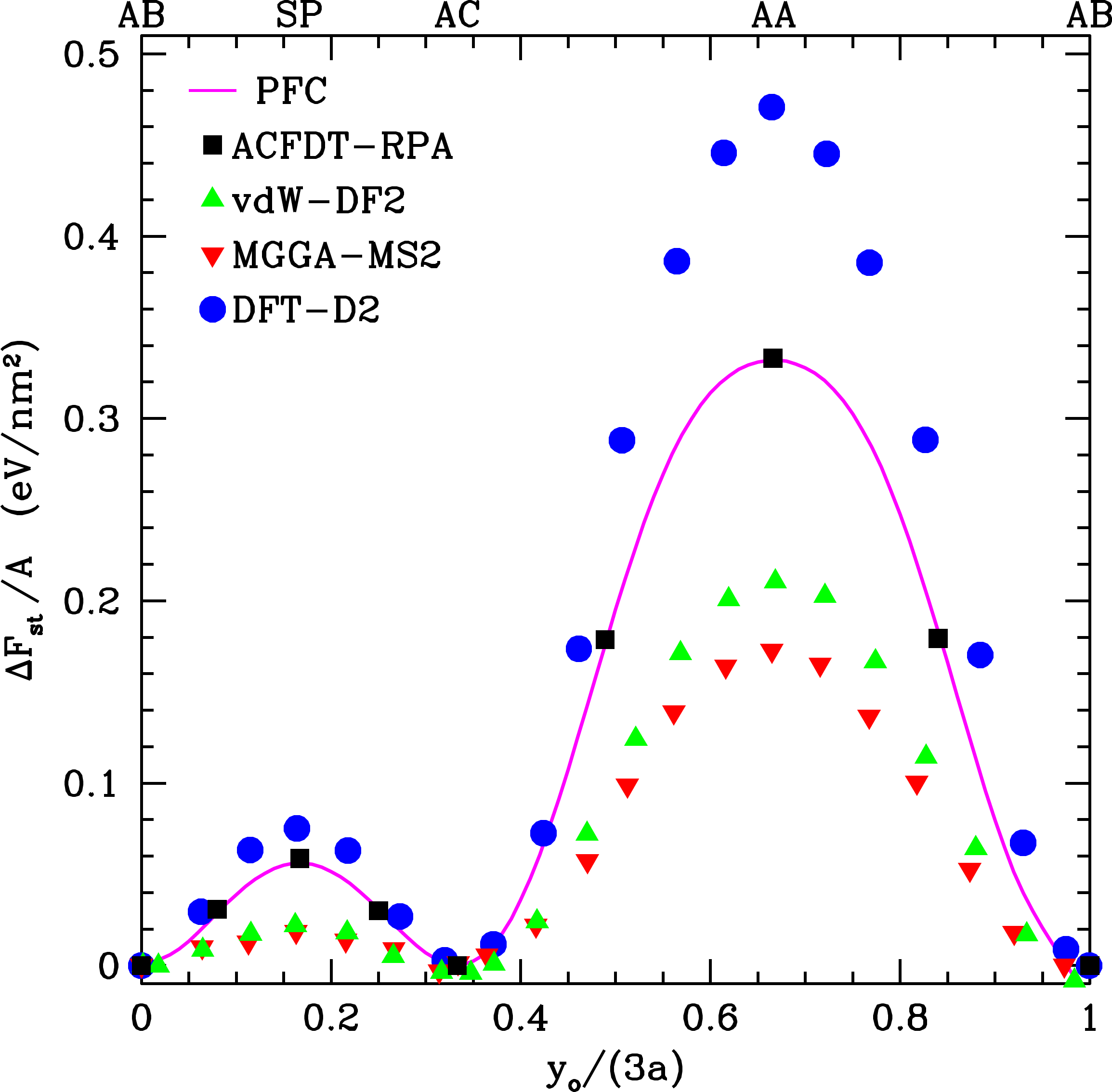} 
\caption{Comparison of stacking free energy difference obtained from various DFT calculations 
\cite{Zhou15} with those from the PFC 
bilayer model. For abbreviations of the different flavors of DFT, see Ref. \cite{Zhou15}}.
\label{fig:FggVc}
\end{figure}

To obtain the best fit to DFT calculations the parameters in Eq.~(\ref{eq:Fc2})
were varied. In the work of Zhou \etal \cite{Zhou15} four different types of DFT were examined. 
Among them three common methods were compared with the most accurate 
and computationally expensive (and consequently more restrictive) method named 
ACFDT-RPA.  Their results showed that these methods generally provided 
qualitatively similar results (except one), which however could be considerably different quantitatively 
(as illustrated in Figs. \ref{fig:HggVc} and \ref{fig:FggVc}).  The 
reader is referred to this reference for more details. For the purpose of this 
work the more accurate and quantitatively reliable ACFDT-RPA results were used to parameterize the PFC model.
Figures \ref{fig:HggVc} and \ref{fig:FggVc}
compare the PFC results against all four DFT calculations of Zhou \etal
for the stacking separation ($\Delta h_{\rm st}$) and free energy
density difference ($\Delta F_{\rm st}/A$ per area $A$) with respect to the AB
stacking, respectively, as a function of shifting distance $y_0$. Table
\ref{tab:gg} summarizes the best fit PFC parameters obtained from the numerical
calculations in these figures.

\begin{table}
\begin{tabular}{|c|c|c|}
\hline
Quantity & Dimensional & Dimensionless \\
\hline
$V^{(1)}_0$ & $1.43$ eV/nm$^2$ & $2.72 \times 10^{-4}$ \\
\hline
$V^{(2)}_0$ & $2.50$ eV/nm$^2$ & $-4.72 \times 10^{-4}$ \\
\hline
$a_2$ & $0.276$ eV/nm$^4$ & $6.51 \times 10^{-5}$ \\
\hline
$\Delta$ & $3.47$ \AA & $10.25$   \\
\hline
$\alpha$ & $-$ & $ 0.26$   \\
\hline
\end{tabular}
\caption{Summary of PFC model parameters for a Gr/Gr bilayer, as identified from the fitting to DFT calculations in Ref.~\cite{Zhou15}.} 
\label{tab:gg}
\end{table}

\subsection{Multiple-Layer Equilibrium}

\begin{figure}[ht]
\vskip 0pt
\includegraphics[width=0.45\textwidth]{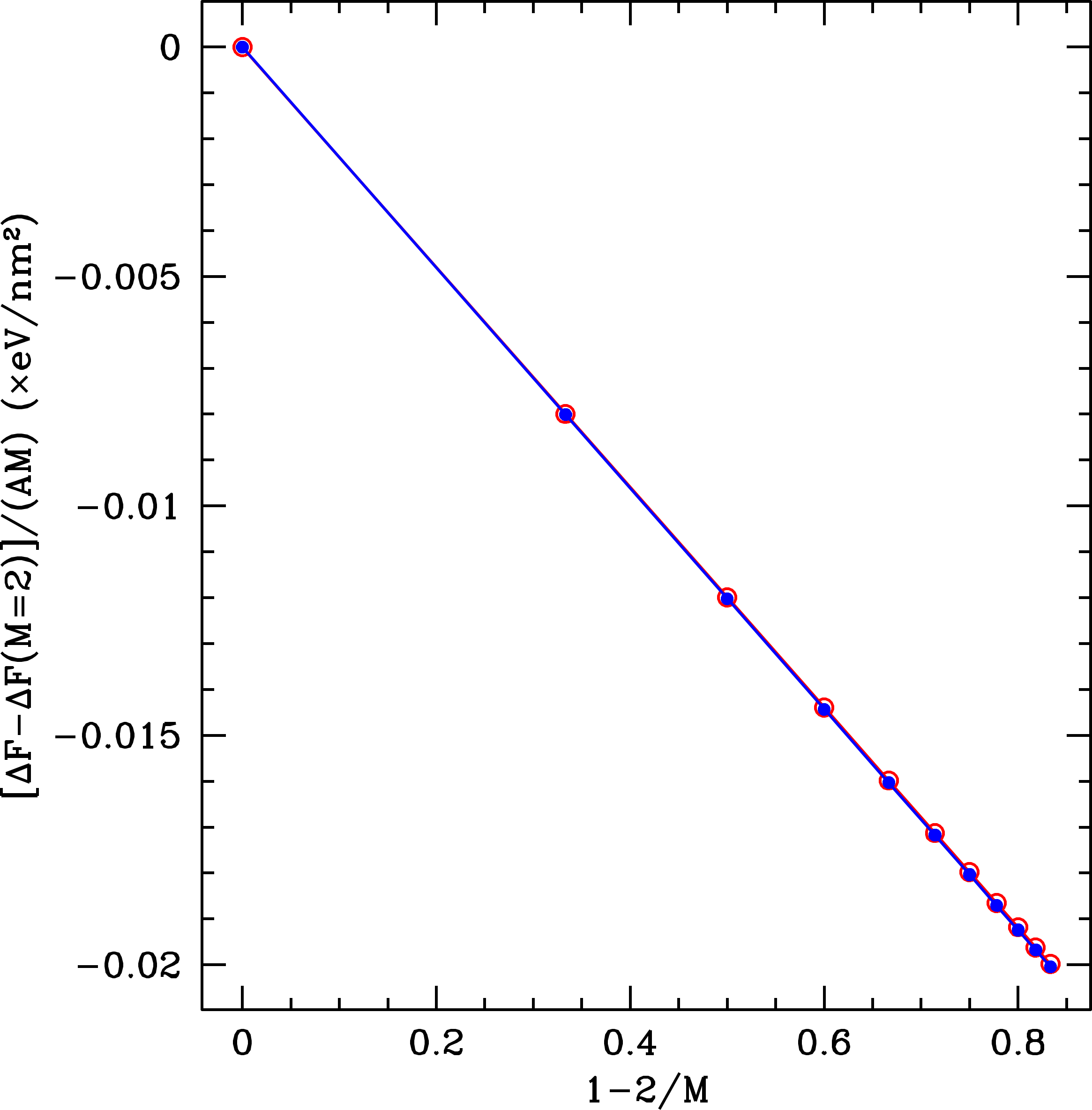} 
\caption{Free energy density difference $\Delta F=F_{\rm t}-2F_{\rm out}$ per layer as a function of $1-2/M$ with number of layers $M$. 
The solid blue (empty red) points are for ABA (ABC) stacking.}
\label{fig:FlayerI}
\end{figure}

In this multilayer PFC model the layers are coupled only to their nearest
neighbours.  This implies that the free energy per unit area per layer for $M$
layers can be written as 
\be \frac{{ F}_{\rm t}}{AM} = \frac{2F_{\rm out}+(M-2)F_{\rm {in}}}{AM},
\ee 
where $F_{\rm out}$ and $F_{\rm {in}}$ are the free energies of the two outer and $M-2$
inner layers respectively.  Thus, 
\be \frac{\Delta F}{AM}=
\left(1-\frac{2}{M}\right)\frac{F_{\rm {in}}}{A}, \label{eq:FAN} 
\ee 
where $\Delta
F={ F}_{\rm t}-2F_{\rm out}$.  This was tested numerically, with results shown in
Fig.~\ref{fig:FlayerI} which verifies Eq.~(\ref{eq:FAN}).   In principle the
layers could stack via an ABA or ABC sequence; however, the free energy
difference between the stackings was found to be very similar, with the ABA
stacking always of slightly smaller energy as shown in Fig.~\ref{fig:FlayerI}. A
linear fitting to these data gives the free energy density per layer $F_{\rm
in}/A = -2.405  \times 10^{-2}$ eV/nm$^2$ for the ABA stacking and $F_{\rm
in}/A=-2.397 \times 10^{-2}$ eV/nm$^2$ for the ABC stacking.

\section{Dislocation pairs in one layer of stacks}
\label{sec:disl}

In a prior study \cite{Elder21} the energy of a pair of dislocations 
in a single monolayer of graphene was examined.  It was found that allowing 
for out-of-plane deformations drastically lowered the energy of 
the pair as compared to a vertically rigid layer.  More specifically, the 
energy of the dislocation pair diverges with system size in the rigid case 
but converges in the flexible case. In what follows an 
examination of how dislocations in one layer influence the 
energetics of a multilayer system of graphene is presented. In principle the addition of layers 
should effectively increase the bending energy coefficient, i.e., the 
additional layers will suppress the bending in the defective layer, leading to a higher 
effective energy of the defects.

\begin{figure}[ht]
\vskip 0pt
\includegraphics[width=0.48\textwidth]{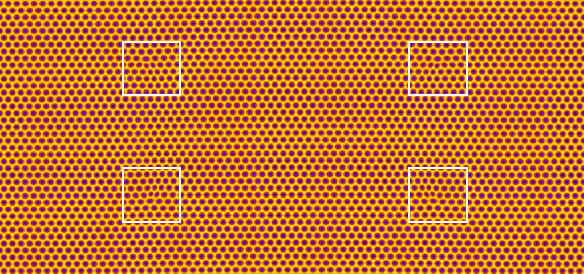} 
\caption{Typical density-field ($n_i$) configuration for $N=69$ which corresponds to a 
system of size $34.0 \times 17.0$ nm$^2$. The boxed areas each contain one 
dislocation.}
\label{fig:defect2}
\end{figure}

\begin{figure}[ht]
\vskip 0pt
\includegraphics[width=0.48\textwidth]{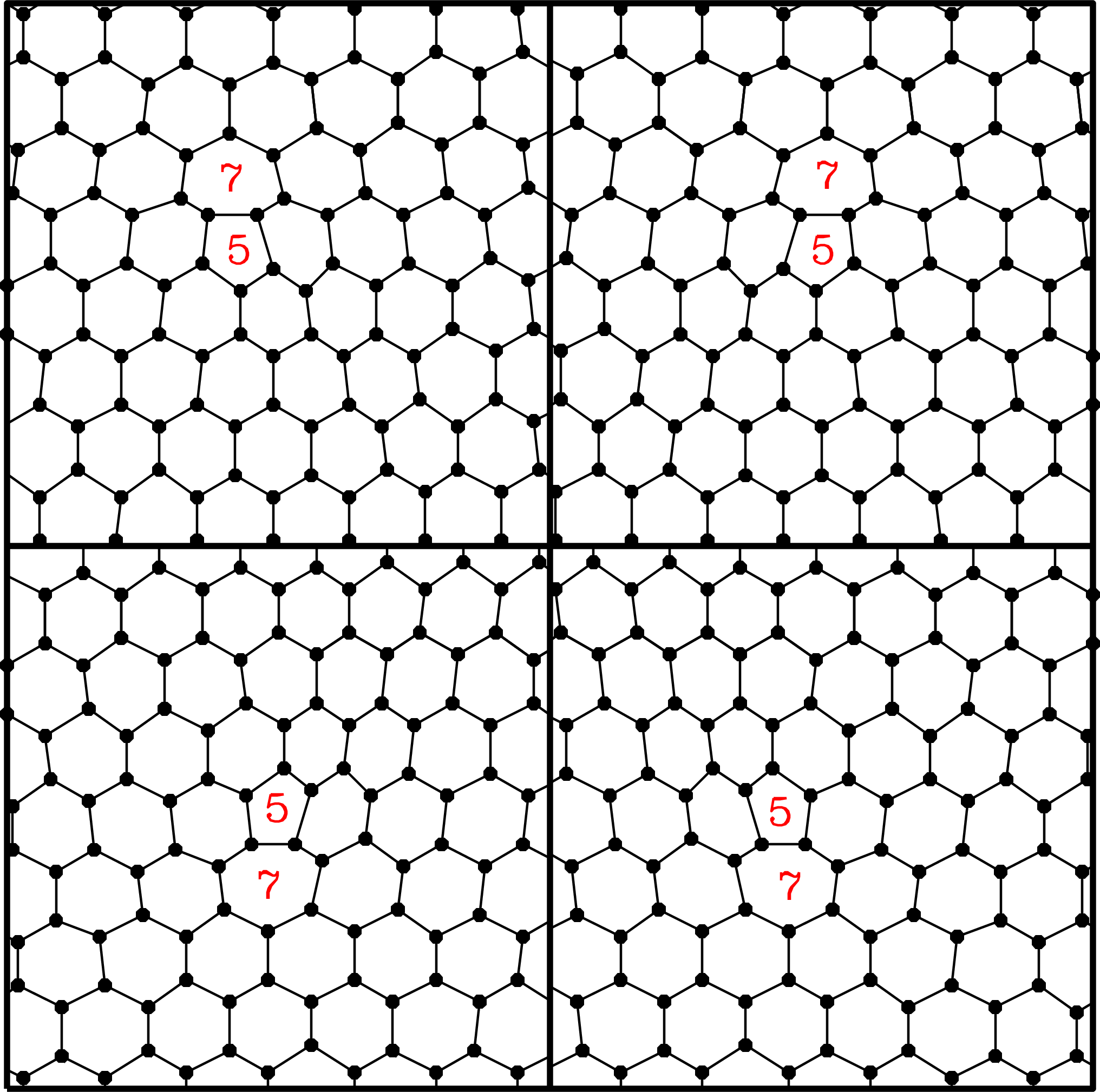} 
\caption{Reconstruction of density maxima for the boxed regions in Fig.~\ref{fig:defect2}, showing 
the classical $5|7$ dislocation structures.}
\label{fig:dotsNC}
\end{figure}

In previous work on a single defected layer, the initial condition was set with
the use of Eq.~(\ref{eq:onem}) for $n_i$ such that in the top (bottom) half of
the 2D system $q_x$ was chosen to fit exactly $N+1$ ($N$) maxima in $n_i$ and
$q_y$ was set to $\qeq$ everywhere.  In addition, a layer of liquid
($n_i=\bar{n}_i)$ of width $20\Delta x$ was inserted at the boundaries
separating top and bottom halves of crystalline domains.  Evolving via
Eqs.~(\ref{eq:ndt}) and (\ref{eq:hdt}) leads to the formation of a pair of
dislocations, i.e., a dislocation dipole. However, this configuration is not
feasible when coupling to other layers that are in equilibrium (such that
$q_x=q_y=\qeq$). For the current simulations the simulation box was set periodic
along both $x$ and $y$ directions and of size $L_x \Delta x \times L_y\Delta y$,
where $L_x$ and $L_y=L_x/2$ are integers. In the non-defected layers the system
size was set to be $L_x\Delta x= 2Na_x$, where $a_x=4\pi/(\sqrt{3}\qeq)$ and $N$
is an integer.  In the defected layer, the $q_x$ value of the top (bottom) half
of the layer was chosen such that there were $2N+1$ ($2N-1$) maxima in the $x$
direction.  The previous single-layer defected configuration could not be used
since it would require $N+1/2$ maxima in the non-defected layers to obtain a
perfect equilibrium state. The configuration used here then leads to two
dislocation dipoles.  Figure \ref{fig:defect2} illustrates a typical
distribution of the density field $n_i$ containing the two dipole pairs, while in
Fig.~\ref{fig:dotsNC} detailed structures of the density maxima and bonds
between neighbouring maxima are shown for the four boxed regions of defects in
Fig.~\ref{fig:defect2}.

One complication of such studies (single- or multi-layer) is that, 
in the layer that contains the defects, the initial condition gives rise 
to a system in which half is under tensile strain and 
the other half in compressive strain.  More specifically this 
leads to a strain $\varepsilon$ in the layer containing the defects,
as given by
\be
\varepsilon= \left\{
\begin{array}{cc}
-1/(2N+1), & \text{if } y>L_y/2, \\
1/(2N-1), & \text{if } y  < L_y/2, 
\end{array}
\right.
\ee
i.e., compressive (tensile) in the top (bottom) half of the layer.
Thus to evaluate the energy of 
the dislocations it must be compared to the energy of strained systems.
For example, if the free energy of the tensile and compressive regions 
is $F_{\rm ten}$ and $F_{\rm com}$ respectively, then the free energy due to the defects
($F_{\rm d}$) is 
\be
F_{\rm d} = { F}_{\rm t}-(F_{\rm ten}+F_{\rm com})/2,
\label{eq:Fd}
\ee
where ${ F}_{\rm t}$ is the total free energy of the entire system given by 
Eq. (\ref{eq:f_tot}).

\subsection{Characterization of strained systems}

In order to evaluate the free energy of defected systems 
(i.e., Eq.~(\ref{eq:Fd})), ${ F}_{\rm t}$ and $F^{\rm c}$ must 
be calculated numerically.  Since height variations in 
the tensile case do not lower the system free energy, 
in an initial setup of this system deviations of the heights from 
their equilibrium values can be set to zero, so that the relaxation to 
equilibrium is relatively rapid.  
In the compressive case it is 
useful to approximate the strain and height of the layers as 
a function of system size and number of layers, and then 
incorporate them in the initial condition.

To gain insight into the properties of the multilayer system it is useful to 
consider the elastic energy of the system in the small deformation limit. 
In a prior work \cite{Elder21} it was shown that in the small deformation limit 
the elastic contribution due to a single layer,  $F_{\rm elas}^1$,  can 
be derived from Eq. (\ref{eq:ftot})  by setting 
\be
n_i = \bar{n}_i+2\phi_{\rm eq}\sum_{j} \cos[\vec G_j\cdot(\vec{r}-\vec{u})],
\ee
where $\vec{u}$ is the displacement field that enters continuum elasticity 
theory.  In the limit of small $|\Grad \vec{u}|$ and $|\Grad h|$ one then gives
\be
\frac{{F}^1_{\rm elas}}{c_{\rm g}} &=& \int d\vec{r}\, \biggl \{ B^x\phi_{\rm eq}^2 \biggl [\frac{9}{2}\left(U_{xx}^2+U_{yy}^2\right)
+3U_{xx}U_{yy} \nline 
&&  \qquad\quad +6 U_{xy}^2\biggr ] + \frac{\kappa}{2} |\Lap h|^2\biggr \},
\label{eq:felas}
\ee
where the strain tensor components, $U_{ij}$, are defined as
\be
U_{\mu\nu}=\frac{1}{2}\left(\partial_\mu u_\nu + \partial_\nu u_\mu + h_\mu h_\nu\right),
\label{eq:strain}
\ee
where $\mu (\nu) = x {\ \rm or \ }y$.
These one-mode approximation results, Eqs.~(\ref{eq:felas}) and (\ref{eq:strain}) for single-sheet elastic free energy, are consistent 
with the well-known form used in the study of flexible sheets \cite{Nelson87} and graphene monolayers \cite{Los09}. 

For a single graphene layer with compressive strain $\varepsilon <0$, 
the height of the film could be written as
\be
h\approx H\cos(Qx), \label{eq:h}
\ee
with the corresponding elastic energy ($F^1_{\rm elas}$ for a single layer) per unit length, 
\begin{equation}
\frac{F^1_{\rm elas}}{L_x c_{\rm g}}=\alpha \varepsilon^2+\frac{1}{2}\left(\frac{1}{2}\kappa Q^2+\alpha 
\varepsilon \right) (QH)^2 + \frac{\alpha}{16}(QH)^4,
\end{equation}
where $Q=2\pi/L_x$, $\alpha=9B^x\phi_{\rm eq}^2/2$, 
and $\phi_{\rm eq}$ is the equilibrium amplitude in the one-mode approximation given 
by Eq. (\ref{eq:phieq}).
In this calculation the elastic free energy has been minimized with respect to the strain, 
resulting in a displacement field 
\be
u_x = \varepsilon x + H^2Q\cos(2Qx)/8.
\ee

In a multiple-layer system within which only one layer is strained, to the lowest-order approximation
\be
\Delta h_{ij}=h_i-h_j=\Delta h_{ij}^0,
\ee
which then gives for $M$ layers,
\be
\frac{F^M_{\rm elas}}{L_x c_{\rm g}}&=&
\frac{F^1_{\rm elas}}{L_x c_{\rm g}}+(M-1)\left[\frac{1}{4}\kappa Q^2 (QH)^2 + \frac{\alpha}{16}(QH)^4\right] \nline
&=&
\alpha \varepsilon^2+\frac{1}{2}\left[\alpha 
\varepsilon+ \frac{MQ^2}{2}\left(\kappa+\frac{\alpha}{8}H^2\right)\right](QH)^2,
\label{eq:Fcomp}
\ee
for the compressive case. 
The free energy of the tensile case is just Eq.~(\ref{eq:Fcomp}) with 
$H=0$, i.e., 
\be
F^M_{\rm elas}/(L_x c_{\rm g})= \alpha \varepsilon^2,
\label{eq:Ften}
\ee
where it has been explicitly assumed that all layers buckle with the same amount, 
which would be valid in the limit of large $a_2$.  Minimizing with respect to 
$H$ then gives
\be
H^2=-2\left(\frac{2\varepsilon}{MQ^2}+\frac{\kappa}{\alpha} \right).
\label{eq:Hp}
\ee
This equation only has a solution if (for compressive strain with $\varepsilon <0$) 
\be
\frac{|\varepsilon|}{Q^2} > \frac{M\kappa}{2\alpha},
\label{eq:epslim}
\ee
naturally showing that if the number of 
layers increases it becomes more  difficult to bend the multilayer 
structure.

Substituting Eq. (\ref{eq:Hp}) into Eq. (\ref{eq:Fcomp}) gives
\be
\frac{F_{\rm elas}^M}{L_x c_{\rm g}}=\frac{M-1}{M}\alpha\varepsilon^2-
\kappa Q^2\left(\varepsilon+ M \frac{\kappa Q^2}{4\alpha}\right),
\label{eq:Fcom}
\ee
which is valid only if Eq. (\ref{eq:epslim}) is satisfied; otherwise the solution 
reduces to Eq. (\ref{eq:Ften}).

\begin{figure}[ht]
\vskip 0pt
\includegraphics[width=0.45\textwidth]{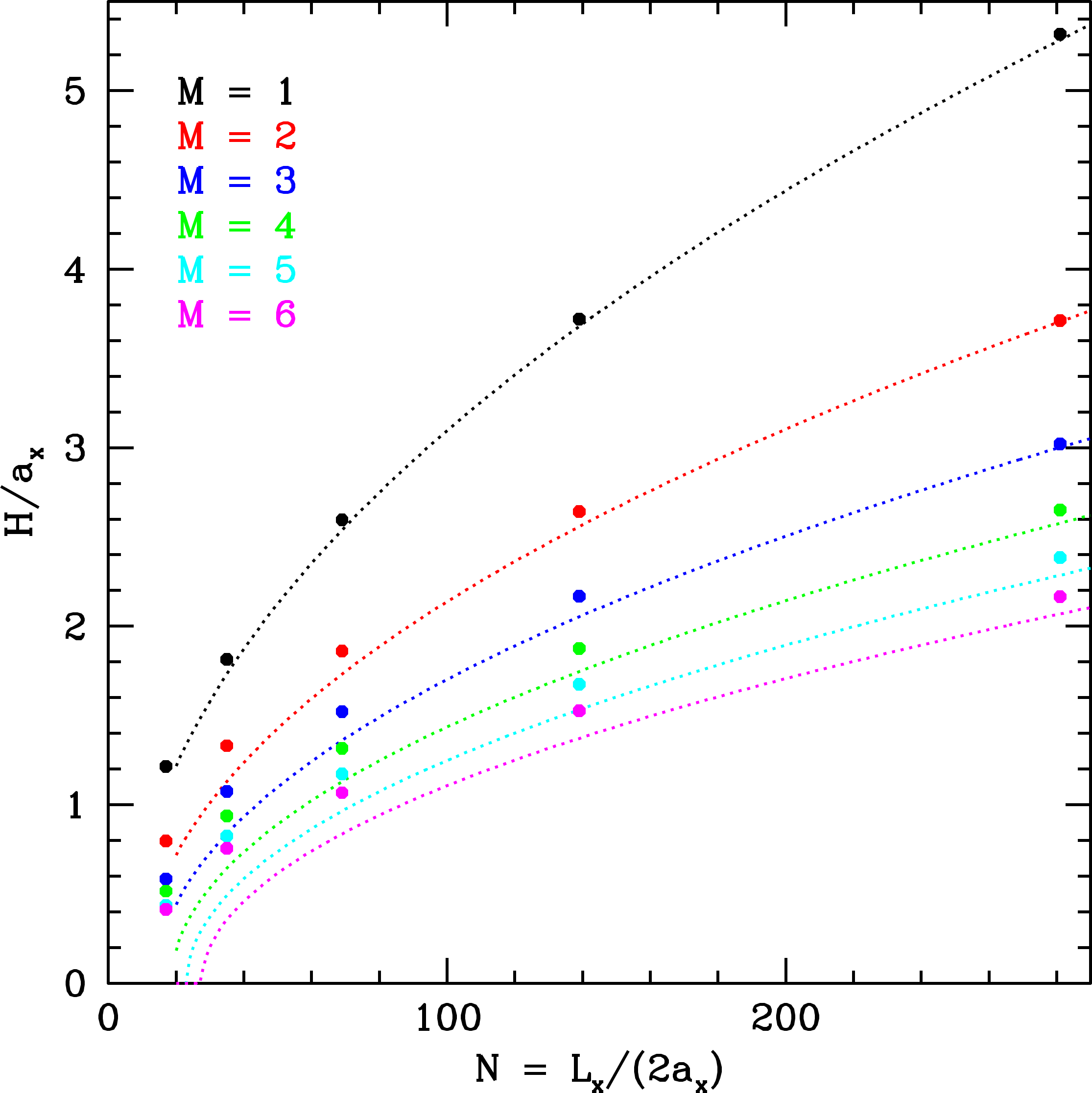}
\caption{Height variation amplitude as a function of system size for different number of layers in the compressive case. Symbols represent the results obtained from numerical simulations of 
the full model, and lines are evaluated from Eq.~(\ref{eq:Hp}).}
\label{fig:Hk1p}
\end{figure}

\begin{figure}[ht]
\vskip 0pt
\includegraphics[width=0.45\textwidth]{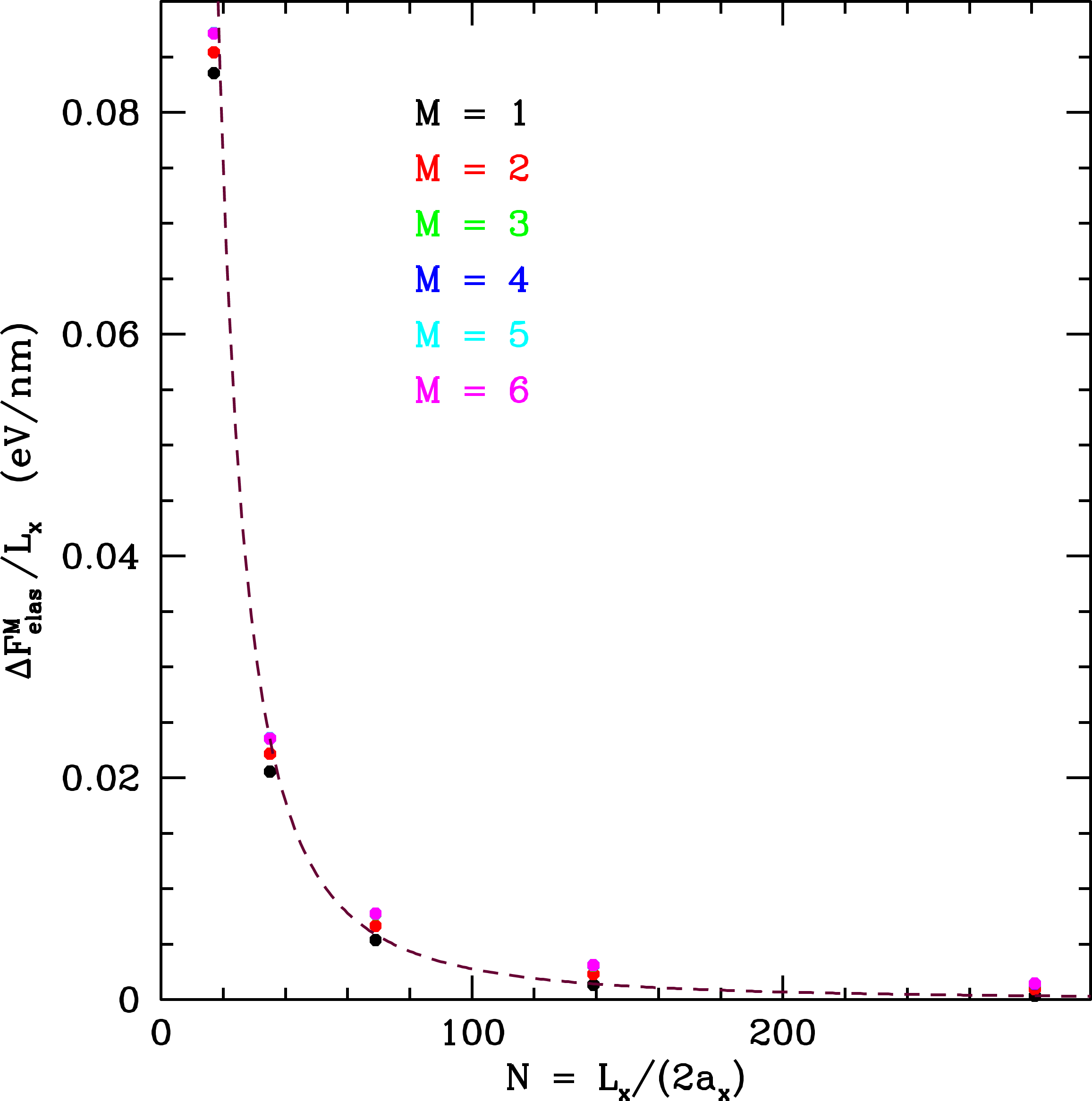} 
\caption{Elastic free energy per length as a function of system size for different number of layers in the tensile case. Symbols represent the results obtained from numerical simulations of 
the full model, and lines are evaluated from Eq.~(\ref{eq:Ften}). 
}
\label{fig:SStrm}
\end{figure}

\begin{figure}[ht]
\vskip 0pt
\includegraphics[width=0.45\textwidth]{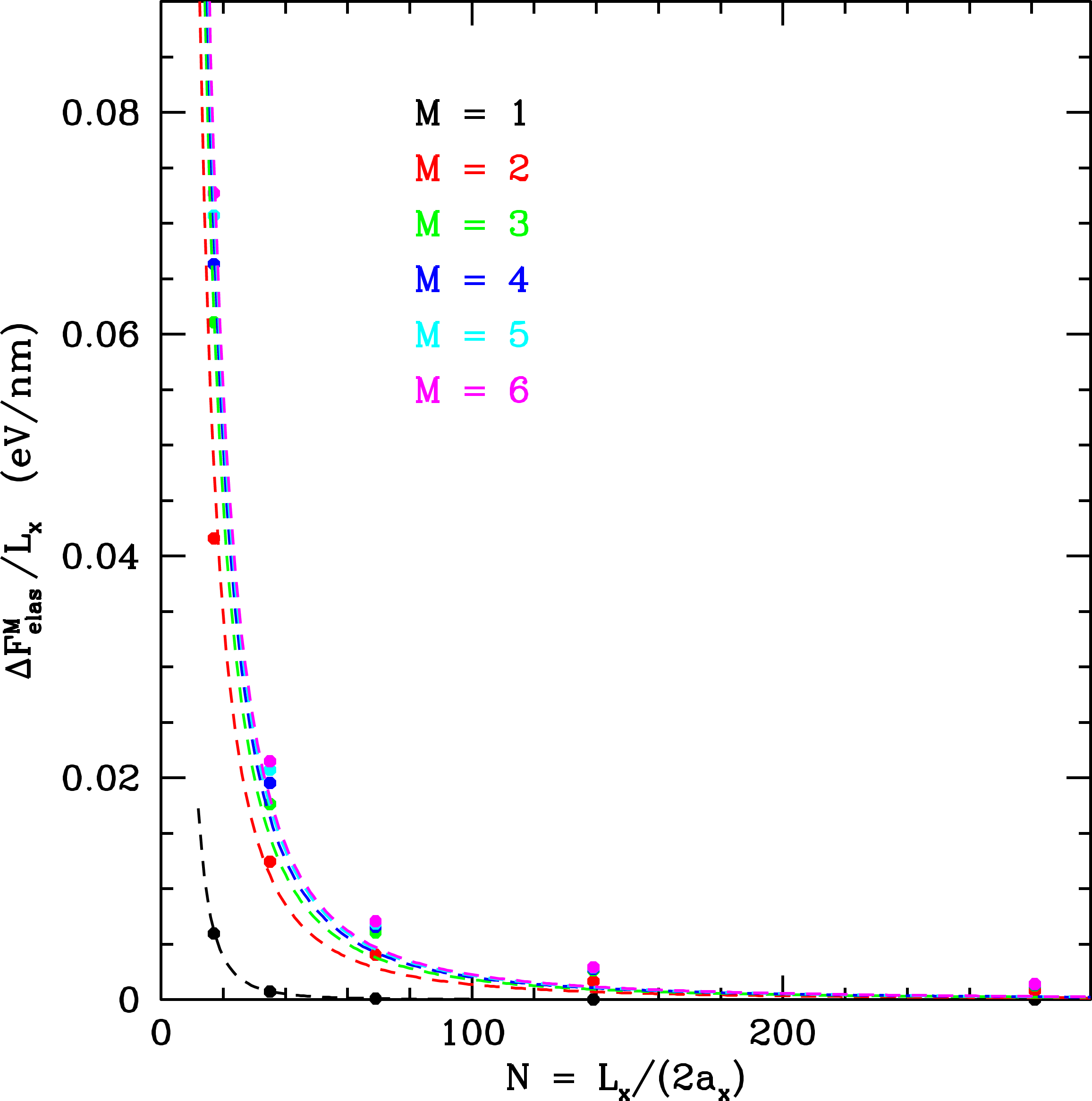} 
\caption{Elastic free energy per length as a function of system size for different number of layers in the compressive case. Symbols represent the results obtained from numerical simulations of 
the full model, and lines are evaluated from Eq.~(\ref{eq:Fcom}). 
} 
\label{fig:SStrp}
\end{figure}

A comparison of this prediction with direct numerical simulations of 
the full model is shown in Fig.~\ref{fig:Hk1p} for $H$ (for the compressive case), and
in Figs.~\ref{fig:SStrm} and \ref{fig:SStrp} for $F_{\rm elas}^M/L_x$ 
(for the tensile and compressive cases respectively).  These figures indicate that 
the above approximate analytic results generally agree with simulation results.
The one anomaly is for $H$ at a system size of $N= 17$ which corresponds to a 
relatively large strain of $5.56$\% and $6.25$\% in the compressive and tensile 
regions respectively.  It should also be noted that the full-order profiles of height
$h$, obtained from the numerical data, indicate the contributions from higher-order modes beyond the approximation of Eq.~(\ref{eq:h}).
Nevertheless the close agreement between the analytic and numerical 
results validate the calculations.

\subsection{Dislocation dipole pair energy}

In this section the influence of the number of layers ($M$) 
and system size ($N$) in a multilayer stacked system which contains one  
defected layer is examined. The defected layer was placed as close 
as possible to the middle of the stack, i.e., directly in the 
middle for an odd number of layers or one more layer on one 
side for an even numbers of layers.  The other layers were 
initialized as perfect crystalline states in an AB stacking 
when possible. While it would be never possible to know if the 
state that emerges from a numerical simulation is the lowest 
possible energy state, several approaches were taken to identify this state.

\begin{figure}[ht]
\includegraphics[width=0.48\textwidth]{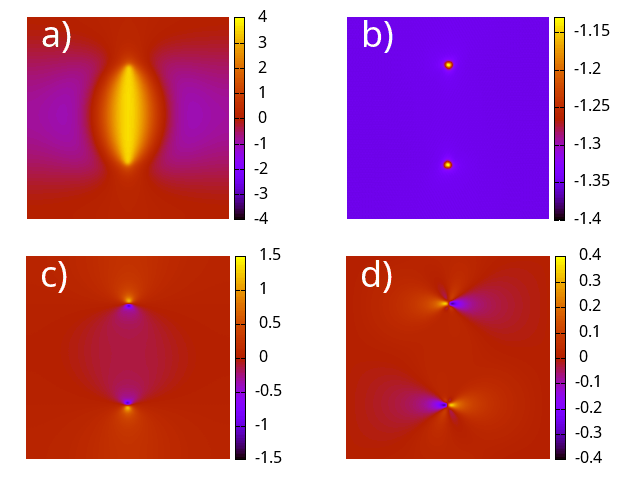} 
\caption{Illustration of Type I structure in a 
bilayer system with $N=139$ ($L_x=69$ nm), for the spatial profiles of (a) height (color-bar scale in \AA), (b) free energy density, 
(c) volumetric stress  $\sigma_{xx}+\sigma_{yy}$, 
and (d) shear stress $\sigma_{xy}$. In (b)--(d) 
the color-bar scale is eV/\AA$^2$. Note that only half of the 
simulation box is depicted here, and (a), (c), and (d) 
are for the defected layer while b) is for the free energy density of 
the total system.}
\label{fig:SJDpa}
\end{figure}

\begin{figure}[ht]
\includegraphics[width=0.48\textwidth]{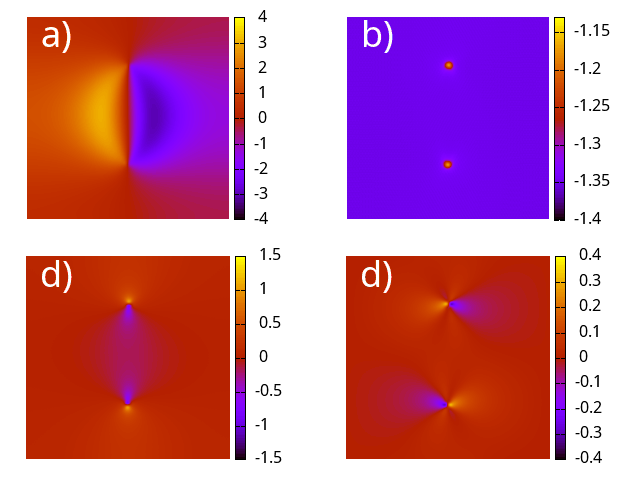} 
\caption{Illustration of Type II structure in a 
bilayer system with $N=139$ ($L_x=69$) nm.  The quantities shown in panels (a)-(d) are the same as those in Fig.~\ref{fig:SJDpa}, with the 
same color-bar scales.}
\label{fig:GSJDpb}
\end{figure}

The first approach was to relax one single layer containing the defects and then
add another layer, and relax the system; then add one more layer and relax again
until a total of five layers had been added to the original defected layer.  As
stated earlier the defected layer was always placed in the middle of the stack
(or closest in the case of an even number of layers).   The second method was to
relax each stack independently (e.g., not rely on using a $M=2$ stack as an
initial condition for a $M=3$ stack).  These approaches led to two distinct
configurations which will be termed Type I and Type II. Figure \ref{fig:SJDpa}
illustrates a Type I configuration in which the height increases at both defect
pairs.  In contrast, in a Type II configuration (Fig.~\ref{fig:GSJDpb}) the
height is positive on one side (in the $x$ direction) of the defect pair and
negative on the other side.   

Which configuration gives the lowest energy is determined by the competition
between local dislocation core energy and elastic strain energy in the
surrounding layers.  Typically the Type II configuration leads to a lower core
energy and a larger strain energy. For the example given in Figs.
\ref{fig:SJDpa} and \ref{fig:GSJDpb} with $(M,N)=(2,139)$, the average core
energy density (defined as the average energy density in a radius of $3.4$ \AA\
around the core) was $-1.1533$ eV/\AA$^2$ for the Type I configuration and
$-1.1544$ for the Type II case. The average volumetric stress
(defined as
$\sigma_V =\sigma_{xx}+\sigma_{yy}$) was $0.0891$ eV/\AA$^2$ for Type I and
$0.1056$ eV/\AA$^2$ for Type II.  In this example the Type II configuration had
the lowest total free energy.

\begin{figure}[ht]
\includegraphics[width=0.48\textwidth]{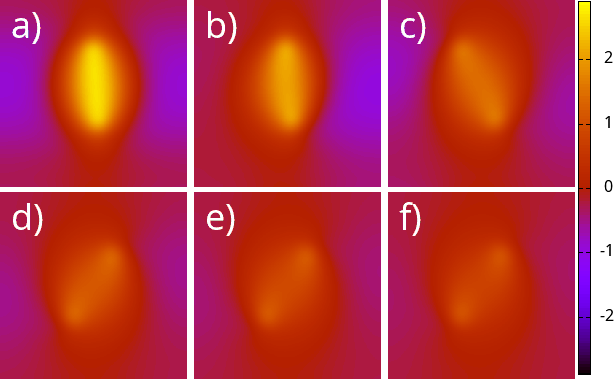} 
\caption{Height profile of the layer containing dislocations with $N=35$ ($L_x=17$ nm), 
corresponding to various multilayer systems with $M=1$ to 6 layers in (a)-(f) respectively.  Only half of the 
system is shown in each panel.
The color-bar scale is in \AA\ and the average height of the layer has been subtracted.}
\label{fig:allh35}
\end{figure}

\begin{figure}[ht]
\includegraphics[width=0.48\textwidth]{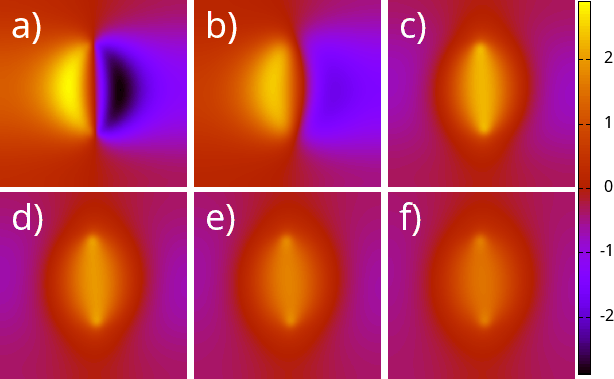} 
\caption{Height profile of the layer containing dislocations with $N=69$ ($L_x=34$ nm) and the setup similar to Fig.~\ref{fig:allh35}, for $M=1$ to 6 layers in (a)--(f) respectively.}
\label{fig:allh69h}
\end{figure}

\begin{figure}[ht]
\includegraphics[width=0.48\textwidth]{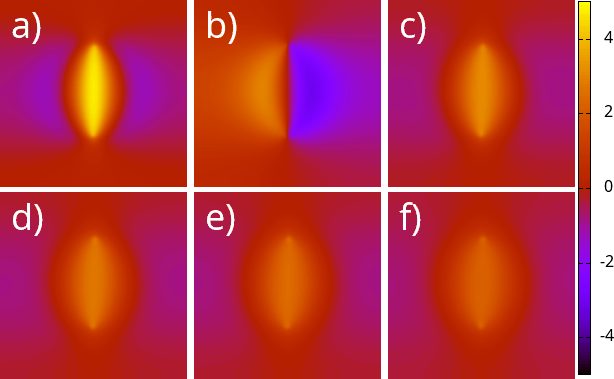} 
\caption{Height profile of the layer containing dislocations with $N=139$ ($L_x=69$ nm) and the setup similar to Fig.~\ref{fig:allh35}, for $M=1$ to 6 layers in (a)--(f) respectively.} 
\label{fig:allh139}
\end{figure}

\begin{figure}[ht]
\includegraphics[width=0.48\textwidth]{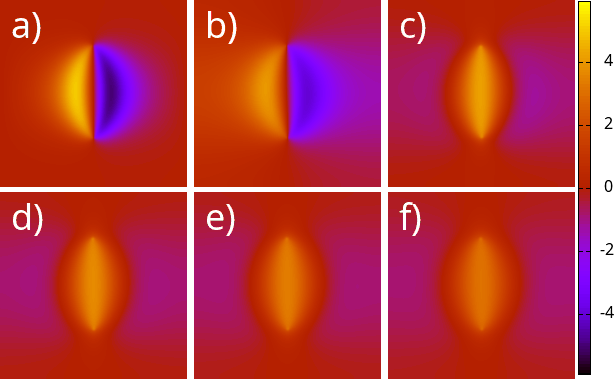} 
\caption{Height profile of the layer containing dislocations with $N=281$ ($L_x=138$ nm) and the setup similar to Fig.~\ref{fig:allh35}, for $M=1$ to 6 layers in (a)--(f) respectively.} 
\label{fig:allh281}
\end{figure}

Figures \ref{fig:allh35}--\ref{fig:allh281} depict the configurations that
minimized the system free energy for system sizes ranging from $N=35$ to
$N=281$. To further check that the lowest energy states were found, initial
condition incorporating final Type I or II configurations were considered.  For
example, a layer could be added to Fig.~\ref{fig:allh139}(b) producing Type II
initial condition for three-layer systems, or a layer could be stripped from
Fig.~\ref{fig:allh139}(c), producing a Type I initial condition for a two-layer
system. Although it was computationally expensive to check the energy of every
system simulated, care has been taken to sample a number of different states.
At the end it became clear that the Type I configuration was the lowest energy
state when the multi-layer stack consists of more than three layers.  
This is due to the fact that more elastic energy is added with each additional layer, since
the defected layer causes out-of-plane fluctuations in other layers. Thus as the
number of layers increases the elastic strain energy dominates over the
dislocation core energy (which exists only in the one defected layer), for which
the Type I configuration exhibits lower strain energy density.  
In addition, for $M>3$, initial conditions containing Type II defects 
would eventually decay into Type I defects.  However, for 
small $M$ the energy difference between Type I and Type II defects was 
very small, indicating that thermal fluctuations may play an important role for $M\le 3$.

\begin{figure}[ht]
\includegraphics[width=0.45\textwidth]{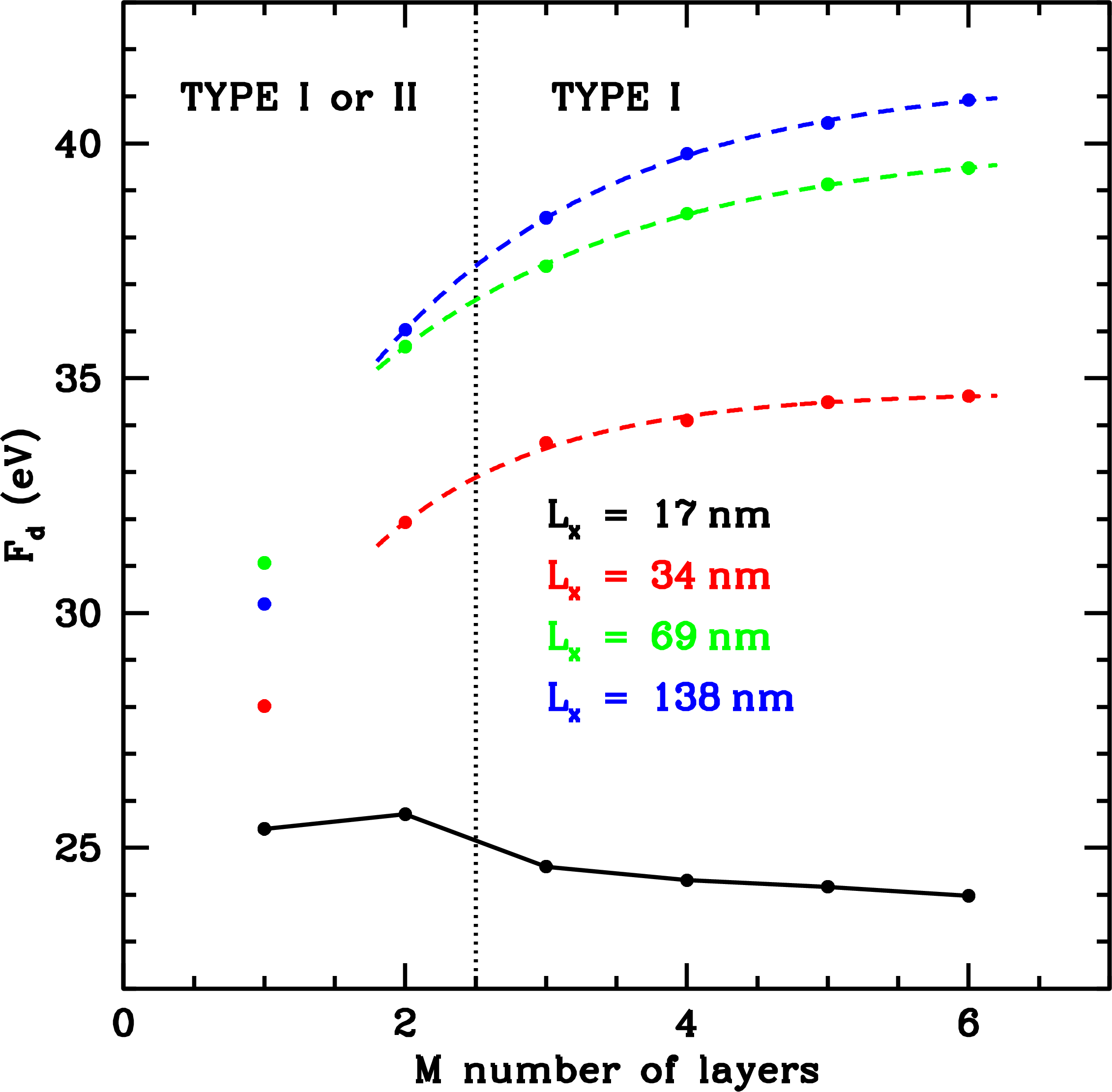} 
\caption{The free energy difference of multilayer system (with two dislocation 
pairs) as a function of the number of layers ($M$) 
for various system sizes. The red, green, and blue dashed 
lines are fits to the functional form 
$F_{\rm d} =  F_{\rm d}^\infty
\left[1-{\rm exp}(-aM-b)\right]$, where $F_{\rm d}^\infty$, $a$, and $b$ are fitting parameters. 
}
\label{fig:dpalln}
\end{figure}

The defect free energy $F_{\rm d}$, defined by Eq.~(\ref{eq:Fd}), for all systems studied 
is summarized in Fig.~\ref{fig:dpalln}. 
As can be seen in 
this figure, for large enough system sizes the defect free energy increases rapidly as the number of 
layers is added up to three layers and then rises at a slower 
rate before starting to saturate.  The one exception was at the smallest 
system examined ($N=35$ with $L_x=17$ nm).  Inspection of the corresponding
configurations (Fig.~\ref{fig:allh35}) reveals strong 
finite size effects.  It is clear that in this case the dipoles (particularly for $M \geq 3$)
tilt away from the vertical $y$ orientation to lower the energy. This also occurred 
in the next biggest system size ($N=69$ with $L_x=34$ nm) but to a 
much less extent, while this tilting phenomenon is not apparent in the two other largest systems 
examined.  

For the three largest systems simulated $F_{\rm d}$ was fit to 
the form $F_{\rm d} = F_{\rm d}^\infty \left[1-{\rm exp}(-aM-b)\right]$, 
where $F_{\rm d}^\infty$, $a$, and $b$ are fitting parameters that can be used to estimate the 
value of $\Delta F_\infty = \Delta F(M\rightarrow \infty)$ which is
the energy difference between ideal bulk and that with one defected layer. These fits (shown in Fig.~\ref{fig:dpalln}) give $\Delta F_\infty=37.71$ eV, $40.04$ eV, 
and $41.44$ eV for $L_x=17$ nm, $34$ nm, and $138$ nm respectively.  This implies 
that the maximum increases of defect energy (i.e., at the $M=\infty$ limit) are given by
$F_{\rm d}^\infty/F_{\rm d}(M=1)=1.24$, $1.29$, and $1.34$ for $L_x=17$ nm, $34$ nm, and 
$138$ nm respectively.  This fit 
also gives an estimate of how many layers it takes to almost reach the 
saturate value of infinite $M$.  From the fitting results, it was found that the number of layers 
needed for $\Delta F_{\rm d}$ to reach 99\% of $\Delta F_\infty$ was approximately 
$M=6$, $10$, and $9$ for $L_x=34$ nm, $69$ nm, and $138$ nm respectively.   

\begin{figure}[ht]
\includegraphics[width=0.48\textwidth]{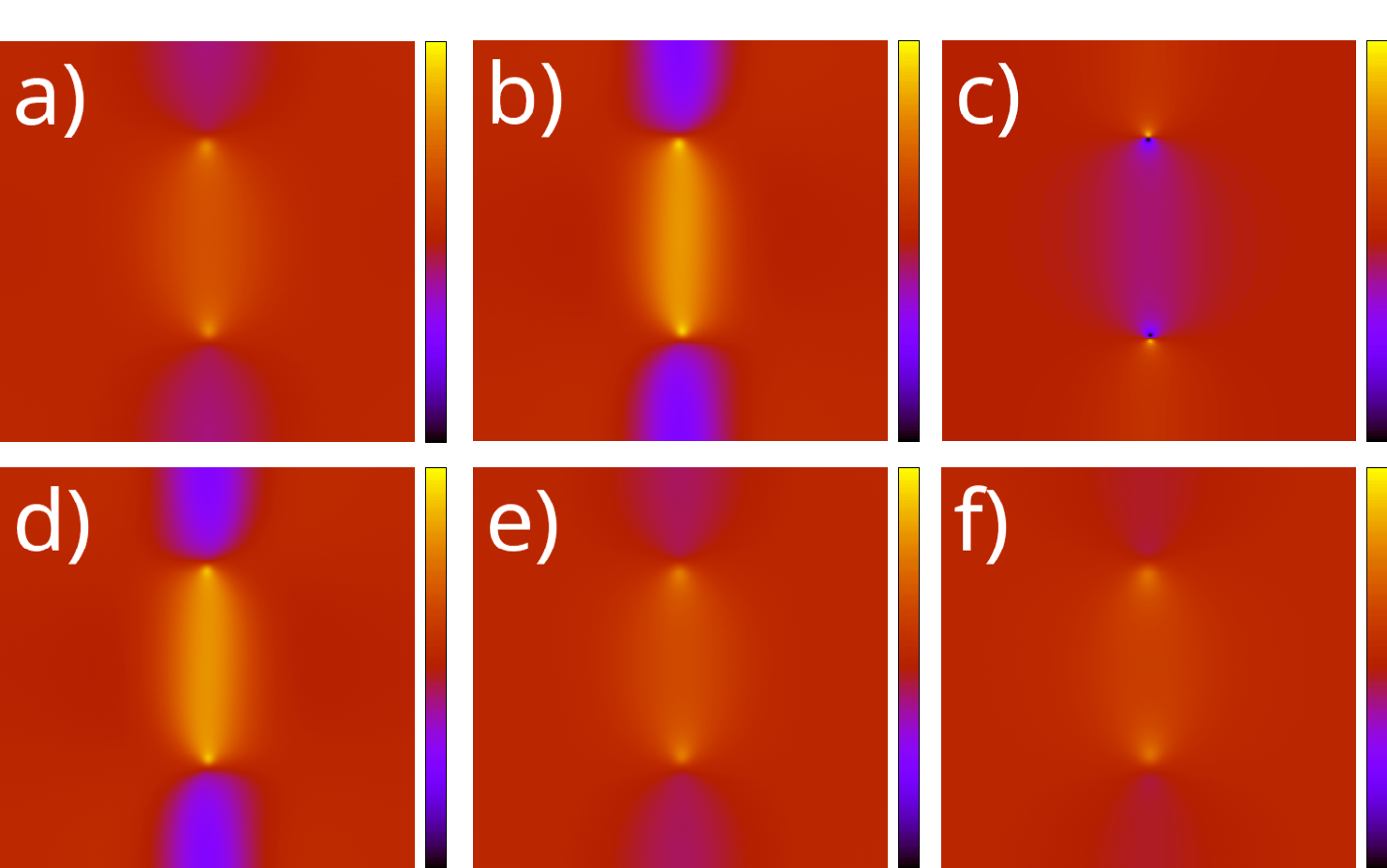} 
\caption{Volumetric stress for a six-layer stack with $N=281$ ($L_x=138$ nm). 
(a)-(f) correspond to layer one to six respectively (where layer three in (c) is the defected layer).  
The color-bar scales in (a), (b), (d)--(f) are from
$-0.1$ eV/nm$^2$ to 0.1 eV/nm$^2$, and in (c) from $-1.5$ eV/nm$^2$ to 1.5 ev/nm$^2$. Note that only 
half of the simulated system is shown here.
}
\label{fig:six_281_sVol}
\end{figure}

\begin{figure}[ht]
\includegraphics[width=0.48\textwidth]{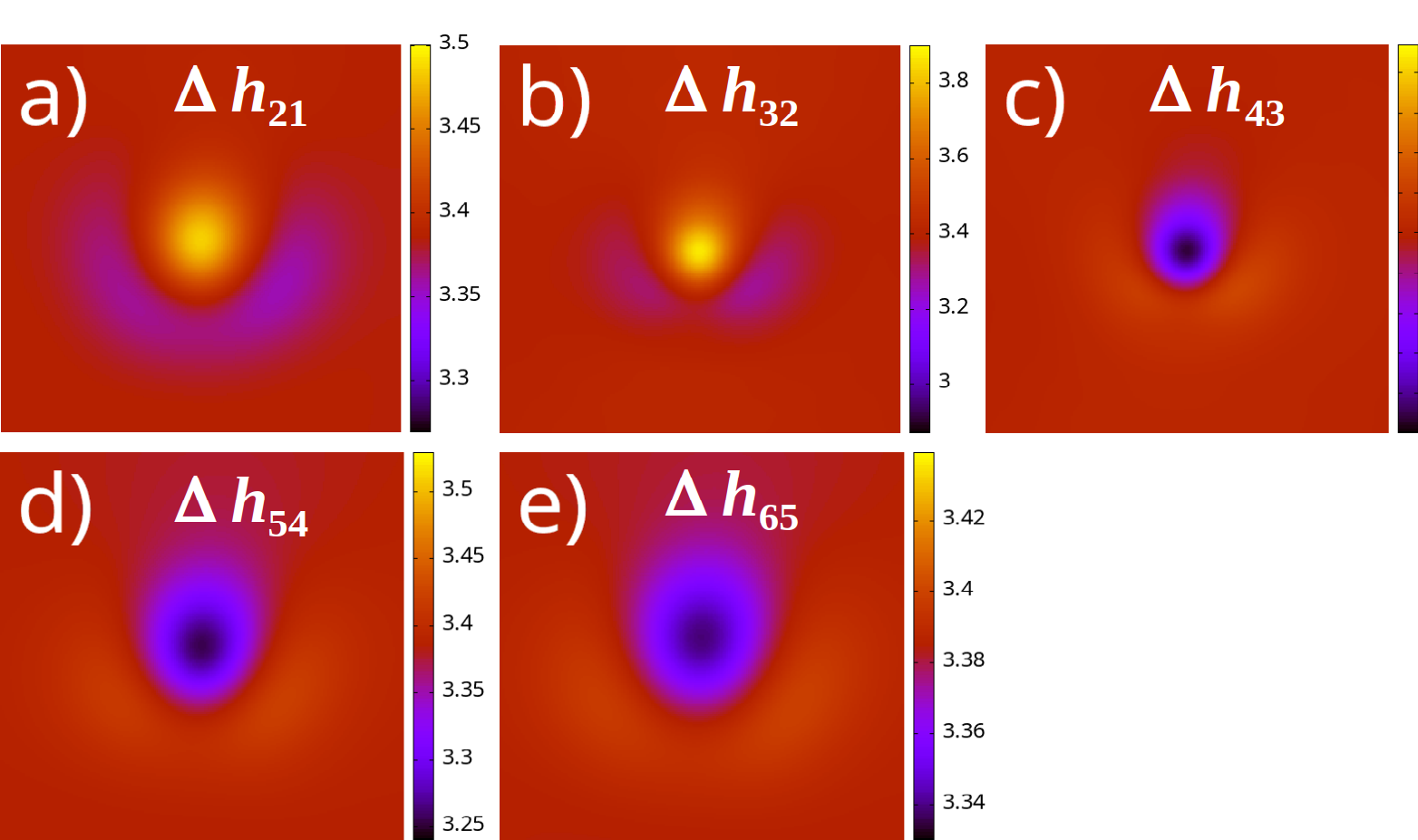} 
\caption{Height difference ($\Delta h_{ij}=h_i-h_j$) between layers around 
one dislocation core (of size 6.8 nm $\times$ 6.68 nm) 
for a stack of six layers with $N=281$ ($L_x=138$ nm). }
\label{fig:six_281_dhZ}
\end{figure}

To further explore the influence of the defected layer on the other 
layers, the volumetric stress, $\sigma_V = \sigma_{xx}+\sigma_{yy}$, 
is shown in Fig.~\ref{fig:six_281_sVol} for each layer of a $M=6$ stack with $N=281$. 
This figure shows how stress in the defected layer is mirrored in the surrounding 
layers, although with smaller magnitudes (by roughly a factor of fifteen).  The 
height difference ($\Delta h_{ij}=h_i-h_j$) near one of the dislocation 
cores is illustrated for 
this same system in Fig.~\ref{fig:six_281_dhZ}, which is constructed such that the 
color-bar scale is $\Delta h_{\rm AB}\pm \Delta$ with $\Delta$ depending on the 
layer differences and $\Delta h_{\rm AB}=3.3845$ \AA\ the equilibrium height 
difference for a AB stacking.  
With this scale the height difference of $\Delta h_{\rm AB}$ appears red in the figures. 
This figure indicates that $\Delta h_{ij}$ is larger below the 
defected layer and smaller above.  This can be understood by considering that the 
height of the defected layer is higher near the dislocation core, causing the height 
of all the layers to increase near the core.  However, since this induces strain,  
the layer heights do not increase enough to reach $\Delta h_{\rm AB}$, 
which leads to height differences above (below) the defected layer to be 
smaller (larger) than the value of $\Delta h_{\rm AB}$.

\section{Discussion and Summary} 
\label{sec:dissum}

In this work the influence of four $5|7$ dislocations (i.e., two dislocation dipole pairs)
in one layer within an otherwise perfect stack of graphene layers has been 
examined.  The 
parameters entering the PFC model were matched to the stacking heights and 
energies calculated by DFT (given in Ref. \cite{Zhou15}) 
to provide quantitative predictions for the system.  Care was taken to 
find the equilibrium states with one layer strained  (compressive or 
tensile) in the stack to provide a reference state to compare with stacks 
containing one layer with the $5|7$ dislocations.

When the initial condition was such that four dislocation defects emerged, it 
was found that the defected layer had a strong impact on the other surrounding
layers as the out-of-plane deformation created by the dislocations 
in the ``middle" layer led to deformations in the surrounding layers, 
which effectively increases the elastic energy of the system.  Two different solutions (Type I and Type II) of multilayer configurations have been identified, 
one with a lower dislocation core energy (Type II) and the other with a lower 
far field elastic energy (Type I).  As the number of layers in the stack was 
increased the Type I configuration was always of lowest energy 
for all the system sizes considered, which is consistent with the fact that far field elastic 
energy in the stack increases with the number of layers added.

Quantitative predictions have been made for the free energy difference, $F_{\rm d}$, of 
a stack of Gr layers containing one defected layer compared to the 
defect-free system.  In general it was found 
that $F_{\rm d}$ increased with system size as was seen in the case of single sheets \cite{Elder21}.  In addition, $F_{\rm d}$ increased with the number of layers ($M$) and 
appears to saturate at some finite value of $M^s \approx 10$.   To some extent
$M^s$ can be considered as the distance over which the defected layer influences 
the entire system.  In addition, for the system sizes studied here our calculations showed that 
inclusion of the four dislocation defects in one embedded layer increased defect energy 
by a maximum of 34\% in the largest film studied.

\begin{acknowledgments}
K.R.E. acknowledges support from the National Science Foundation (NSF) under
Grant No. DMR-2006456, Oakland University Technology Services and
high-performance research computing facility (Matilda), and
Aalto Science Institute for support during his visit. Z.-F.H. acknowledges
support from NSF under Grant No. DMR-2006446. T.A-N. has been supported in 
part by the Academy of Finland through its Quantum Technology Finland CoE 
Grant No. 312298 and the European Union -- NextGenerationEU instrument Grant No. 353298. 
The authors wish to acknowledge
CSC -- IT Center for Science, Finland, for generous computational resources and
the CSC HPC support group for helping setup and run the  PFC solver.   The authors 
would like to thank Cristian Achim for useful discussions and CUDA code development. 
We 
would also like to thank Arsalan Hashemi for useful discussions and carrying out additional DFT calculations for stacked graphene layers.
\end{acknowledgments}

\appendix

\section{Functional Derivatives}
Defining $F^{\rm c}$ as the free energy due to the coupling between layers, i.e., 
\be
F^{\rm c} = \sum_{i=1}^{M-1} {\cal F}^{c}_{i+1,i},
\ee
gives 
\be
\frac{\delta F_{\rm c}}{\delta n_i}&=& V_0(\delta n_{i-1}+\delta n_{i+1})  \nline
&& -2a_2 \Delta \alpha\left[
(\Delta h_{i,i-1}-\Delta h^0_{i,i-1})\delta n_{i-1} \right. \nline 
&& \left. +(\Delta h_{i+1,i}-\Delta h^0_{i+1,i})\delta n_{i+1}
\right], 
\ee
for $i\ne 1$ or $i \ne N$, while for $i=1$,
\be
\frac{\delta F_{\rm c}}{\delta n_1}&=& V_0\delta n_{2}  
 -2a_2 \Delta \alpha
(\Delta h_{2,1}-\Delta h^0_{2,1})\delta n_{2},
\ee
and for $i=N$
\be
\frac{\delta F_{\rm c}}{\delta n_N}&=& V_0\delta n_{N-1}  \nline
 &&-2a_2 \Delta \alpha
(\Delta h_{N,N-1}-\Delta h^0_{N,N-1})\delta n_{N-1}.
\ee

The functional derivatives with respect to height $h_i$ are given by: 
for $i=1$,
\be
\frac{\delta F_{\rm c}}{\delta h_1}
&=&-2a_2\left[
h_2-h_1-\Delta h_{1,2}^0 \right] \nline
&=&2a_2\left[
h_1-h_2+\Delta h_{1,2}^0 \right],
\ee
for $i=N$,
\be
\frac{\delta F_{\rm c}}{\delta h_N}=2a_2\left[
h_N-h_{N-1}-\Delta h_{N,N-1}^0
\right],
\ee
and for $i\ne 1$ or $i \ne N$,
\be
\frac{\delta F_{\rm c}}{\delta h_i}&=&2a_2\left[
h_{i}-h_{i-1}-\Delta h_{i,i-1}^0
-h_{i+1}+h_{i}+\Delta h_{i,i+1}^0
\right] \nline 
&=& 2a_2\left[
2h_{i}-h_{i-1}-h_{i+1}
+\Delta h_{i,i+1}^0
-\Delta h_{i,i-1}^0
\right].
\ee

\bibliography{Nlayers}

\end{document}